\documentclass[letterpaper]{article} 
\usepackage{aaai25}
\usepackage{times}  
\usepackage{helvet}  
\usepackage{courier}  
\usepackage[hyphens]{url}  
\usepackage{graphicx} 
\usepackage{natbib}  
\usepackage{caption} 
\usepackage{algorithm}
\usepackage{algorithmic}

\usepackage{newfloat}
\usepackage{listings}
\DeclareCaptionStyle{ruled}{labelfont=normalfont,labelsep=colon,strut=off} 
\floatstyle{ruled}
\newfloat{listing}{tb}{lst}{}
\floatname{listing}{Listing}
\title{CS-Eval: A Comprehensive Large Language Model Benchmark for CyberSecurity}
\author{
    Zhengmin Yu\equalcontrib \textsuperscript{\rm 1},
    Jiutian Zeng\equalcontrib \textsuperscript{\rm 2},
    Siyi Chen \textsuperscript{\rm 2},
    Wenhan Xu \textsuperscript{\rm 3},
    Dandan Xu \textsuperscript{\rm 3},
    Xiangyu Liu \textsuperscript{\rm 2},
    Zonghao Ying \textsuperscript{\rm 3},
    Nan Wang \textsuperscript{\rm 1},
    Yuan Zhang \textsuperscript{\rm 1},
    Min Yang \textsuperscript{\rm 1}
}

\affiliations {
    \textsuperscript{\rm 1}Fudan University\\
    \textsuperscript{\rm 2}Alibaba Group\\
    \textsuperscript{\rm 3}School of Cyber Security, UCAS\\
}

\usepackage{bibentry}

\usepackage{listings} 

\begin{document}
\maketitle
\begin{abstract}
Over the past year, there has been a notable rise in the use of large language models (LLMs) for academic research and industrial practices within the cybersecurity field. However, it remains a lack of comprehensive and publicly accessible benchmarks to evaluate the performance of LLMs on cybersecurity tasks. To address this gap, we introduce CS-Eval, a publicly accessible, comprehensive and bilingual LLM benchmark specifically designed for cybersecurity. CS-Eval synthesizes the research hotspots from academia and practical applications from industry, curating a diverse set of high-quality questions across 42 categories within cybersecurity, systematically organized into three cognitive levels: knowledge, ability, and application. Through an extensive evaluation of a wide range of LLMs using CS-Eval, we have uncovered valuable insights. For instance, while GPT-4 generally excels overall, other models may outperform it in certain specific subcategories. Additionally, by conducting evaluations over several months, we observed significant improvements in many LLMs' abilities to solve cybersecurity tasks. The benchmarks are now publicly available \footnote{\url{https://github.com/CS-EVAL/CS-Eval}}.
\end{abstract}
\section{Introduction}
Large language models (LLMs) have demonstrated remarkable capabilities across various domains, including finance\cite{kim2024financial,lee2024survey}, healthcare\cite{cascella2023evaluating}, education\cite{latif2023knowledge}, legal analysis\cite{chen2024survey}, and scientific research\cite{sun2024scieval}, significantly enhancing human abilities and productivity. These models can now solve professional problems that were previously inaccessible to non-experts. This new problem-solving paradigm\cite{saha2024llm,yao2024survey} has also emerged in the field of cybersecurity, where LLMs show promising potential. In academia, substantial research is focused on utilizing LLMs for cybersecurity tasks\cite{Examing, sun2024llm4vuln}. In industry, alongside powerful general-purpose LLMs like GPT\cite{GPT-4, Claude}, specialized models such as SecGPT\cite{SecGPT} are being developed to tackle security-specific challenges.

Currently, there are numerous general-purpose benchmarks for large language models, such as MMLU\cite{MMLU} and GLUE\cite{GLUE}. Additionally, in specialized fields like finance and law, there are comprehensive evaluation benchmarks\cite{fineval, fei2023lawbench}. However, the landscape of evaluation benchmarks specifically designed to assess the cybersecurity capabilities of large language models (LLMs) remains relatively underdeveloped. While general benchmarks provide broad evaluations, they often overlook the unique challenges inherent to cybersecurity. Conversely, cybersecurity-specific datasets offer detailed assessments of particular tasks but lack the comprehensive scope needed for a thorough evaluation. This gap underscores the need for a benchmark that integrates both breadth and depth, facilitating a more accurate evaluation of LLMs in cybersecurity. Such a benchmark would not only help developers identify the limitations of their models but also assist users and researchers in selecting and refining the most suitable models for their needs.

To evaluate the cybersecurity capabilities of large language models (LLMs), several challenges must be addressed:
\begin{itemize}
    \item Establishing Clear and Effective Principles for Evaluating the Cybersecurity Capabilities of LLMs:
    Applying LLMs to cybersecurity research involves numerous application scenarios and capabilities, which are continually increasing in number and gaining momentum\cite{yao2024survey}. Ensuring that the evaluation comprehensively covers various fields and cognitive levels is a significant challenge.
    \item Ensuring Benchmark Quality:
    The high quality of evaluation data, including question accuracy and data contamination prevention, is crucial to the reliability of results. In the cybersecurity domain, question design requires technical precision and logical rigor. Additionally, the prior exposure of publicly available cybersecurity data to large models further heightens the risk of data contamination.
    \item Overcoming High-Level Scoring Limitations:
    LLM benchmarks in other fields, like SciEval \cite{SciEval}, typically focus on large-grained scoring dimensions. However, cybersecurity spans diverse specialized domains with unique challenges. Relying solely on high-level scoring can lead to overly abstract evaluations, failing to capture a model’s effectiveness in specific tasks. For example, evaluating a model’s capability in vulnerability detection requires specific evaluation tasks tailored to different types of vulnerabilities. Without this focus, high-level scores may misrepresent LLM’s effectiveness on the specialized domain.
\end{itemize}

To tackle the challenges, we introduce CS-Eval, a comprehensive benchmark designed for a thorough assessment of LLMs' cybersecurity capabilities. CS-Eval's broad and deep evaluation criteria offer a reliable test suite to drive advancements in LLM development within the cybersecurity domain.
Specifically, CS-Eval first define the evaluation principle from multiple perspectives—including academic research, industry expertise, and domain-specific applications—covering 11 categories and 42 subcategories, such as vulnerability management, threat detection, and data security. Next, a high-quality dataset of 4,369 carefully crafted questions, including multiple-choice, true/false, and subjective types, is developed by a team of five experts over one month. This dataset is rigorously designed, validated, and supported by a dynamic update strategy that regularly refreshes the benchmark data, ensuring continuous improvement and mitigating data contamination risks. Lastly, through meticulous task and capability categorization, we deliver detailed evaluations of model performance across various tasks. This approach not only identifies the overall proficiency of the models but also provides a nuanced evaluation of the strengths and weaknesses in different domains, offering valuable insights for future model enhancement and development. 

Through the CS-Eval benchmark evaluation, key trends in LLM performance in cybersecurity were revealed. GPT-4 8K excelled in general tasks, while models like Qwen2-72B-Instruct performed best in specialized domains, emphasizing the potential value of domain-specific optimization. Data quality proved critical, as lower-quality data led to underperformance in domain-specific models like SecGPT-13B. Larger models generally outperformed smaller ones, but efficient designs like MoE models also held their own. Over time, improvements in data quality and training strategies led to smaller models sometimes surpassing larger predecessors. The incorporation of synthetic data further enhanced model capabilities, particularly in specialized fields. Additionally, we provided helpful insights to improve large language models' capabilities in the cybersecurity domain.

\textbf{Contributions.}
This paper makes the following contributions:
\begin{itemize}
\item We introduce CS-Eval, the first open-accessible bilingual comprehensive cybersecurity benchmark that encompasses a broad range of tasks and domains, offering thorough and precise assessments of large language models (LLMs). Our benchmark dataset is publicly accessible at \url{https://github.com/CS-EVAL/CS-Eval}.
\item We address the challenges of creating a cybersecurity benchmark by aligning with industry and academic priorities, ensuring rigorous data quality, and enhancing practical insights. These strategies also provide valuable guidance for developing benchmarks in other specialized fields.
\item Through extensive experiments conducted over various time periods, we obtained several important findings, such as the top-performing models for different tasks and the manifestation of the scaling law in our benchmark. More importantly, we offered practical insights for future domain-specific training of large language models.

\end{itemize}

\begin{figure*}[h]
\centering
\includegraphics[width=0.9\textwidth]{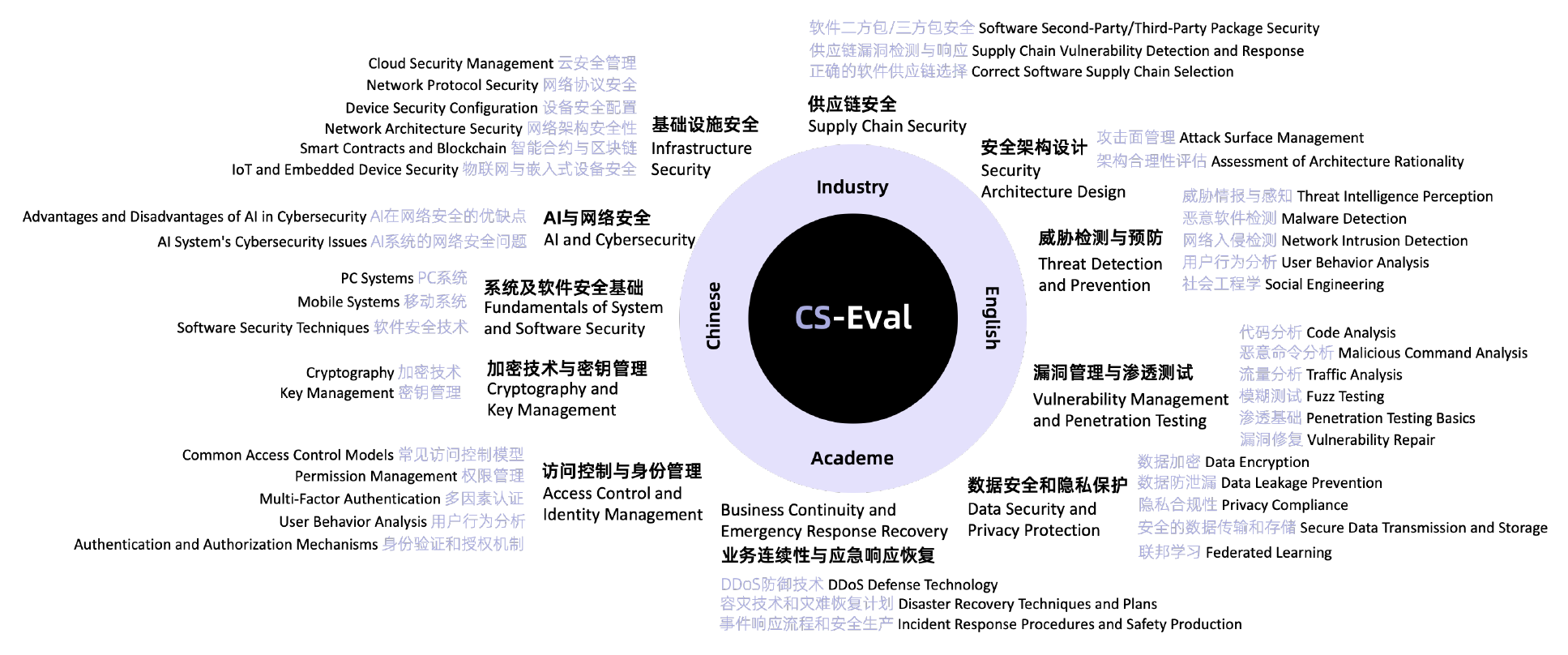}
\caption{Overview of Fields Covered by CS-Eval: A comprehensive cybersecurity benchmark encompassing 11 categories and 42 subcategories across various domains.}
\label{fig:overview}
\end{figure*}

\section{Related Work}
\subsection{LLM Benchmark}
For comprehensive LLM evaluation, several benchmarks have been developed to assess model performance across a diverse range of tasks. The GLUE benchmark \cite{GLUE} was one of the first to establish a multi-task natural language understanding platform, optimizing the evaluation of NLU tasks. SuperGLUE\cite{superglue}, introduced after GLUE, offers more challenging tasks and an enhanced evaluation framework to further test the capabilities of language models. In addition, the MMLU benchmark \cite{MMLU} was introduced to evaluate models across 57 tasks, including mathematics, computer science, and law. The HELM benchmark \cite{HELM} aggregates 42 different scenarios and evaluates LLMs using 7 metrics, ranging from accuracy to robustness, providing a broad assessment of model capabilities. Similarly, Big-Bench \cite{BIG-Bench} includes challenging tasks from multiple domains to gauge task complexity and model capabilities. Additionally, C-Eval\cite{C-Eval} was introduced to assess advanced knowledge and reasoning abilities within the Chinese context, reflecting cognitive evaluation across multilingual and multicultural dimensions. Specific benchmarks like AGIEval\cite{AGIEval} also play a role in evaluating models' performance on standardized tests, focusing on simulating human cognitive abilities.

To evaluate the specific professional abilities of LLMs, specialized benchmarks have been developed to provide a more detailed assessment of their performance across various domains. 
In the medical domain, benchmarks like PubMedQA \cite{PubMedQA} and MedMCQA \cite{Medmcqa} focus on evaluating LLMs' ability to understand and generate responses to medical questions, reflecting their potential utility in healthcare. LegalBench \cite{guha2024legalbench} assesses models on their ability to comprehend and generate legal text, providing insight into their applicability in legal contexts. Fineval \cite{fineval}, on the other hand, evaluates models on tasks specific to the financial domain, such as sentiment analysis and question answering based on financial texts. SciEval further extends this evaluation by testing LLMs on their scientific reasoning and problem-solving capabilities. Recently, SecBench \cite{Secbench} and SecEval \cite{li2023seceval} were introduced to evaluate the cybersecurity capabilities of LLMs. However, they primarily focus on knowledge-based questions, lacking practical, real-world scenarios.

\subsection{CyberSecurity-specfic Dataset}
For specific and granular tasks in cybersecurity, there exist long-established datasets that have been widely used in research. As mentioned in the introduction, these datasets are highly specialized in their evaluation focus. Below, we detail some of the prominent research areas and the datasets associated with these trending topics.
For vulnerability detection, datasets like SySeVR\cite{SySevr} and FLVD\cite{FLVD}, sourced from the National Vulnerability Database (NVD) and GitHub, provide information on vulnerabilities linked to corresponding code. Vulbench\cite{Vulbench} aggregates a diverse range of these datasets, including those from Capture The Flag (CTF) challenges and real-world applications, with annotations detailing the vulnerability type and root cause for each function.
In binary analysis, the GNU Coreutils and BinaryCorp datasets are frequently utilized. The GNU Coreutils dataset includes various Unix/Linux command-line tools, while BinaryCorp contains binaries from ArchLinux repositories compiled at different optimization levels, making them useful for tasks like binary code similarity detection and vulnerability analysis\cite{wang2022jtrans}.
For vulnerability repair, ExtractFix\cite{Extractfix} and GPT2-CSRC\cite{Examing} are notable datasets. ExtractFix guides patch synthesis through semantic reasoning, and GPT2-CSRC, trained on curated C/C++ code, optimizes prompt-based vulnerability repair.
Log analysis research often relies on the Loghub dataset\cite{Loghub}, a large-scale collection of log data from distributed systems, supercomputing environments, mobile systems, and server applications. The ADFA SID datasets\cite{creech2013generation} are also commonly used, focusing on abnormal logs generated through network attacks.
For test case generation, Detect4J\cite{defects4j} and Bugs.jar\cite{saha2018bugs} are widely utilized. Detect4J offers Java vulnerability programs across multiple versions, while Bugs.jar provides a larger and more diverse collection of vulnerabilities.

\section{The CS-Eval Benchmark}

\begin{figure*}[h]
\centering
\includegraphics[width=0.8\textwidth]{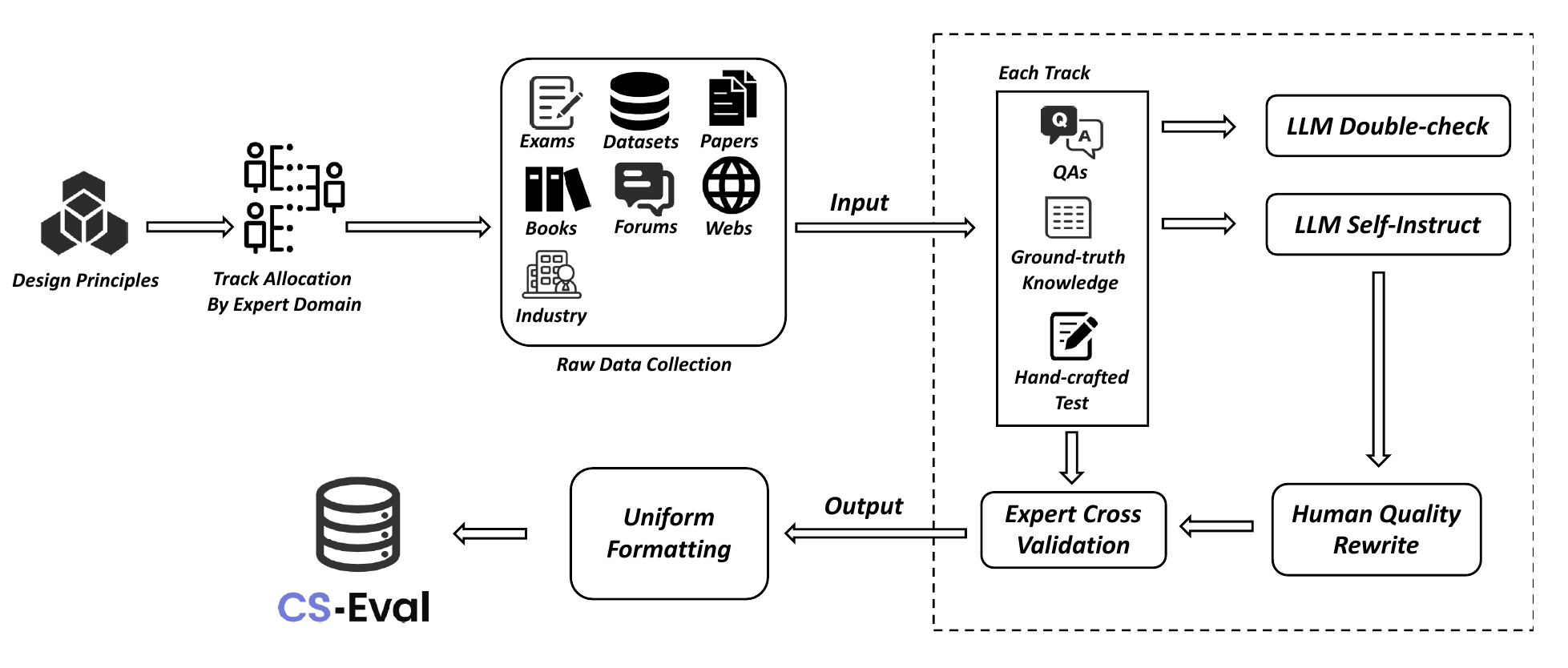}
\caption{CS-Eval Data Collection Process}
\label{fig:data_collection}
\end{figure*}

\subsection{Design Principle}


The design of the CS-Eval dataset is primarily inspired by two key sources: the session structures of top-tier academic conferences in computer security, and practical security practices and benchmark design within the industry.
From an academic perspective, we specifically analyzed the session setups and related paper titles from the "Big Four" cybersecurity conferences \cite{SP2024, NDSS2024, CCS2024, SEC2024} over the past three years, extracting key terms to identify current research hotspots.
In addition, within the industrial context, we analyzed the design principles of numerous other benchmarks, such as those demonstrated in OpenCompass \cite{OpenCompass2024}, which includes 58 large model benchmarks. We also consulted engineers who apply large models to cybersecurity tasks to understand their evaluation needs. This comprehensive approach helped us shape the design principles of CS-Eval, considering both general benchmark frameworks and the practical requirements of the field.
Based on this analysis, we structured the levels and scope necessary for evaluating large models in the field of cybersecurity, identifying three key dimensions: knowledge, capabilities, and application. Specifically:
\begin{itemize}
    \item Knowledge level: This includes 11 major categories and 42 subcategories, covering practical industrial scenarios and the research needs of security professionals, as illustrated in Figure \ref{fig:overview}.
    \item Capabilities level: This focuses on specific skills required to complete various tasks, including comprehension, code analysis, scientific computing, reasoning, long-text processing, and summarization abilities.
    \item Application level: This covers practical application scenarios within the cybersecurity field, such as vulnerability analysis, binary analysis, log analysis, network traffic analysis, penetration testing, digital forensics, malware analysis, and cryptanalysis tasks.
\end{itemize}

CS-Eval has established scoring evaluations for different domains, which can provide feedback to model developers on their data mix, enabling them to achieve optimal data proportions and further realize balanced performance improvements across various tasks. Additionally, we understand that some model developers may focus more on tasks within a specific domain. Therefore, the scores for domains can help model developers gauge model performance, and researchers can use the experimental results from these domains as a reference factor when choosing models for further study.

\subsection{Dataset Overview}

To address the challenges and apply the principles we discussed earlier, we have designed five types of questions: multiple-choice, multiple-answer, true/false, subjective, and experimental. These questions cover three different levels of depth and span 42 categories in breadth. According to the dataset construction process, we first gathered and curated a large collection of static questions. This collection, built and cross-verified by 5 people over the course of a month, represents high-quality data. Then, we leveraged this foundation to continue generating dynamic data, ensuring ongoing improvements and resilience against data contamination and shifts in large model behaviors. This dynamic approach allows us to adapt to changes and maintain the relevance and accuracy of our dataset.

\subsection{Data Collection}
The construction of the CS-Eval dataset as Figure \ref{fig:data_collection} shows that primarily consists of four steps.

\textbf {Step1: Knowledge collection in various domains of cybersecurity.} We first find research and industry experts from different domains, then divide the task to them, ensuring expert and data sources diversity for each sub-domain. Using this method, to avoid potential data limitations. These knowledge sources include the internet, course materials and exam resources from various universities, sanitized adaptations of practical implementations in LLMs, reliable conclusions from relevant domain papers, and cutting-edge research.

\textbf{Step2: Construction of questions.} Initially, we categorize the types of data sources. For data already formatted for exams or Q\&A purposes, we employ human adaptation to formulate questions. For textual knowledge-based data, we design different question formats and construct Self-Instruct\cite{wang2022selfinstruct} templates as outlined in Table \ref{table:instruct}, then generate specific types of questions by embedding this knowledge into Self-Instruct prompts and invoking GPT4.

\textbf{Step3: Question validation and quality enhancement.} The adapted and generated questions undergo quality checks and revisions by humans to avoid issues such as inaccuracies in questions produced by large models, low-quality content, and disproportionate options in multiple-choice questions.

\textbf{Step4: Cross-validation of questions.} Finally, all questions are compiled and subjected to cross-validation to ensure consistency and correctness across the entire dataset.
\subsection{Dynamic Data Generation}
In the CS-Eval dataset, dynamic data generation is employed to maintain continuous data variation, preventing models from artificially inflating their scores by training on test questions. To accomplish this, we utilize capable LLMs like GPT4o and employ dynamic rewriting strategies to regularly generate new data by revising questions across all 11 major categories, drawing from our meticulously curated high-quality datasets.

Two primary strategies are employed. The first strategy involves instructing the model to rewrite questions by applying techniques such as altering the subject, modifying options, and restructuring the logical flow. The second strategy involves providing the model with a set of questions from a specific category, prompting it to identify and summarize the underlying knowledge points, which are then recombined to generate new, original questions. The prompts used for these strategies are detailed in Appendix C\ref{chap:appendix_c}. Additionally, all generated questions will undergo a quick manual review to ensure data accuracy.

\begin{table}[h!]
\centering
\begin{tabular}{p{1\linewidth}}
\hline
\textbf{Top Category:} Vulnerability Management and Penetration Testing \\ \hline
\textbf{Sub Category:} Vulnerability Repair \\ \hline
\textbf{Prompt:} \\ 
Multiple-Choice Question\\
Context:
\begin{lstlisting}[basicstyle=\scriptsize\ttfamily]
type InfoResp struct {
    GroupId int 
    InfoId int 
    InfoStatus int 
    InfoContent string }
\end{lstlisting}
Function:
\begin{lstlisting}[basicstyle=\scriptsize\ttfamily]
func GetInfo(ctx *gin.Context, req *InfoReq) (*InfoResp, error) {
    if req.InfoId == nil {
        logs.CtxError(ctx, "InfoId is nil")
        return nil, common.NewError(common.ReqError, "InfoId is nil")}
\end{lstlisting}
\begin{lstlisting}[basicstyle=\scriptsize\ttfamily, numbers=none]
    [A] infoResp, err := info.GetInfo(req.InfoId)
    [B] if err != nil {
        logs.CtxError(ctx, "Get Info error")
        return nil, err}
    [C] if infoResp == nil {
        logs.CtxError(ctx, "Info is nil")
        return nil, common.NewError(common.RespError, "Info is nil")}
    [D] return infoResp, nil}
\end{lstlisting}
Fix\_Code\_Block:
\begin{lstlisting}[basicstyle=\scriptsize\ttfamily]
if infoResp.GroupId != common.GetGroupID(ctx) {
    logs.CtxError(ctx, "GroupId not match, req:%v, infoResp:%v", req, infoResp)
    return nil, common.NewError(common.RespError, "GroupId not match")}
\end{lstlisting}
Please select the most appropriate position to use 'fix code block' for fixing an authorization bypass vulnerability based on the above function and context.\\ 
A: [A] B: [B] C: [C] D: [D]
\\ \hline
\end{tabular}
\caption{A Multiple-Choice Question Example on Vulnerability Management in CS-Eval}
\label{table:case}
\end{table}
  
\section{Experiment}
 
\subsection{Setup}
We evaluate Large Language Models (LLMs) using the CS-Eval benchmark. Each question within CS-Eval is accompanied by a carefully crafted prompt, as illustrated in Table \ref{table:case}. This example features a multiple-choice question, and the CS-Eval benchmark provides the input prompt. Various LLMs are then employed to generate inferences based on these prompts. After the inference process is completed, the responses are collected and standardized into a unified format to ensure consistency. Subsequently, we calculate the scores using predefined metrics. The following sections provide a detailed overview of our model selection criteria and the evaluation metrics utilized in this study.
\subsubsection{Models.}
To comprehensively assess the cybersecurity capabilities of Large Language Models (LLMs), we selected a range of popular LLMs that are publicly accessible. These models include both open-source and closed-source variants, with parameter sizes ranging from small to large, as shown in Table \ref{table:models}. For all open-source models, we prioritized local deployment. We obtained the model parameters directly from the official sources and maintained them in their original form, avoiding any modifications such as quantization. All configurations and settings, including hyperparameters like inference temperature, were strictly aligned with the official defaults to ensure consistency and reliability in the evaluation.
\begin{table}[t]
\centering
\resizebox{\columnwidth}{!}{
\begin{tabular}{l|l|l|l}
\hline
\textbf{Model} & \textbf{Creator} & \textbf{\#Parameters} & \textbf{Access} \\ \hline
GPT4o & OpenAI & unpublic & API \\
GPT4-8K & OpenAI & unpublic & API \\
GPT3.5-Turbo-16K & OpenAI & unpublic & API \\
DeepSeek-V2-0628 & DeepSeek & 236B & API\\
Qwen-14B-Chat & Alibaba Cloud & 14B & Weights \\
Qwen1.5-14B-Chat & Alibaba Cloud & 14B & Weights \\
Qwen1.5-MoE-A2.7B-Chat & Alibaba Cloud & 14.3B & Weights \\
Qwen2-7B-Instruct & Alibaba Cloud & 7B & Weights\\
Qwen2-72B-Instruct & Alibaba Cloud & 72B & Weights\\
Baichuan-13B-Chat & BaiChuan-Inc & 13B & Weights \\
Baichuan2-13B-Chat & BaiChuan-Inc & 13B & Weights \\
360Zhizhao-7B-Chat-4K & 360 & 7B & Weights \\
Mistral-7B-Instruct-v0.2 & Mistral AI & 7.3B & Weights \\
Yi-6B-Chat & 01.AI & 6B & Weights \\
ChatGLM3-6B & Zhipu AI & 6B & Weights \\
ChatGLM4 & Zhipu AI & 9B & API\\
SecGPT-13B & Clouditera & 13B & Weights \\
Llama-2-13b-chat-hf & Meta & 13B & Weights \\
Llama-3.1-8B-Instruct & Meta & 8B & Weights\\
Llama-3.1-70B-Instruct & Meta & 70B & Weights\\
\hline

\end{tabular}}

\caption{Models Evaluated by CS-Eval: Creator, Parameter Count, and Access Type.}
\label{table:models}
\end{table}

\subsubsection{Metrics.}
For each question within CS-Eval, we employ accuracy as the primary metric, with evaluations tailored to the specific type of question. For multiple-choice and true/false questions, accuracy is assessed by an exact match between the submission and the designated answer. Before comparison, we utilize LLMs combined with regular expressions to extract the relevant answers and format them into a standardized JSON structure to ensure consistency.
For open-ended questions, where responses are more subjective, LLMs play a crucial role in determining correctness. These models assign binary labels (0 or 1) based on the alignment of the output with reference answers, ensuring a thorough and consistent evaluation across all question types.
The final score for each field is calculated as the average score of all questions within that field.

\begin{table*}[]
\centering
\resizebox{\textwidth}{!}{
\begin{tabular}{lccccccc}
\hline
\multicolumn{1}{c}{Model} & \multicolumn{1}{c}{Business Continuity and} & \multicolumn{1}{c}{Security Architecture} & \multicolumn{1}{c}{Supply Chain} & \multicolumn{1}{c}{Data Security and} & \multicolumn{1}{c}{Threat Detection} & \multicolumn{1}{c}{Vulnerability Management} & \multicolumn{1}{c}{\textbf{Average}} \\
\multicolumn{1}{c}{} & \multicolumn{1}{c}{Emergency Response Recovery} & \multicolumn{1}{c}{Design} & \multicolumn{1}{c}{Security} & \multicolumn{1}{c}{Privacy Protection} & \multicolumn{1}{c}{and Prevention} & \multicolumn{1}{c}{and Penetration Testing} & \multicolumn{1}{c}{\textbf{Score}} \\
\hline
GPT4-8K                  & 84.28        & 83.90  & 89.30 & 86.90     & 85.21   & 89.63     & \textbf{87.57}   \\
Qwen2-72B-Instruct       & 77.33        & 85.37  & 86.38 & 86.31     & 88.56   & 88.19     & \textbf{86.82}   \\
GPT4o                    & 77.00        & 83.41  & 89.70 & 84.92     & 87.82   & 85.44     & \textbf{86.11}   \\
DeepSeek-V2-0628         & 78.33        & 83.90  & 89.37 & 83.33     & 84.87   & 85.60     & \textbf{85.26}   \\
ChatGLM4                 & 79.67        & 82.44  & 90.70 & 83.73     & 84.13   & 85.44     & \textbf{84.99}   \\
Llama-3.1-70B-Instruct   & 74.33        & 79.02  & 86.71 & 86.11     & 85.61   & 84.95     & \textbf{84.32}   \\
Qwen2-7B-Instruct        & 73.00        & 80.00  & 87.38 & 79.56     & 84.32   & 82.52     & \textbf{82.63}   \\
GPT3.5-Turbo-16K         & 81.27        & 76.59  & 88.96 & 82.14     & 79.52   & 80.71     & \textbf{80.59}   \\
Qwen-14B-Chat            & 78.60        & 74.15  & 87.63 & 76.68     & 79.67   & 77.80     & \textbf{79.04}   \\
Llama-3.1-8B-Instruct    & 71.67        & 76.59  & 85.05 & 75.99     & 76.57   & 77.99     & \textbf{77.34}  \\
Qwen1.5-14B-Chat         & 70.23        & 70.73  & 81.27 & 77.58     & 77.53   & 75.77     & \textbf{76.66}   \\
Qwen1.5-MoE-A2.7B-Chat   & 72.24        & 68.78  & 81.94 & 70.24     & 76.61   & 74.80     & \textbf{74.63}   \\
Baichuan2-13B-Chat       & 73.91        & 71.71  & 80.27 & 75.69     & 76.94   & 70.55     & \textbf{73.92}   \\
Baichuan-13B-Chat        & 69.00        & 70.73  & 75.42 & 71.43     & 69.37   & 71.04     & \textbf{70.73}   \\
360Zhinao-7B-Chat-4K     & 66.33        & 68.78  & 70.00 & 65.02     & 68.63   & 64.78     & \textbf{66.37}   \\
Mistral-7B-Instruct-v0.2 & 63.67        & 62.44  & 72.76 & 63.44     & 64.40   & 63.71     & \textbf{65.93}   \\
Yi-6B-Chat               & 59.67        & 60.00  & 72.76 & 62.85     & 63.85   & 64.68     & \textbf{65.27}   \\
ChatGLM3-6B              & 56.67        & 61.46  & 68.44 & 57.71     & 61.47   & 50.81     & \textbf{57.33}   \\
SecGPT-13B               & 45.33        & 45.85  & 59.14 & 43.08     & 47.34   & 46.77     & \textbf{47.34}   \\
Llama-2-13b-chat-hf      & 39.13        & 37.07  & 30.43 & 35.52     & 39.00   & 38.57     & \textbf{38.08} \\
\hline
\end{tabular}}
\caption{Parts of Evaluation Results}
\label{table:eval1}
\end{table*}

\subsection{Overall Comparison}
The overall experimental results are partially presented in Table \ref{table:eval1}. We report both the average scores and the scores for certain specific fields. Detailed scores for each field are provided in Table \ref{table:all_combined} in the Appendix.

Across all fields, OpenAI's GPT-4 8K leads with an impressive average score of 87.57, demonstrating superior performance across various cybersecurity domains. Despite the introduction of newer versions like GPT-4o, GPT-4 8K remains the top performer, showcasing its robust general capabilities and consistent excellence in cybersecurity tasks. The performance difference can likely be attributed to GPT-4o's optimization for multimodal capabilities, enhanced speed, and efficiency, which make it versatile but sometimes less focused on text-only tasks compared to GPT-4 8K. Following closely, Qwen2-72B-Instruct also achieves a high average score of 86.82, indicating strong performance in specialized tasks. Other models, such as DeepSeek-V2-0628 and ChatGLM4, also score competitively, with averages of 85.26 and 84.99, respectively.

The score range of the evaluation results, spanning from 87.57 to 38.08, highlights significant variation in performance across different models. This variation reflects substantial differences in the models' capabilities for specific tasks in cybersecurity. These disparities can likely be attributed to several factors, including variations in model architecture, differences in training data types and quality, and optimization for specific domains. Interestingly, although general-purpose models like GPT-4 perform well on average, other models may excel in specific areas. This variability may stem from the specialized optimization of different models based on their training tasks and data, enabling them to perform better in certain tasks. For example, the Qwen2-72B-Instruct model excels in areas such as Threat Detection and Prevention, with a score of 88.56, surpassing GPT-4's score of 85.21 in the same category.

\subsection{Domain-Specific LLM's Behavior}
Domain-specific LLMs \cite{roziere2023codellama} are typically further trained from large volumes of domain-specific professional data based on general-purpose LLMs, enabling them to leverage the unique value of domain data. Generally, domain models face the challenge of improving performance on domain knowledge and tasks while seeing a decline in general capabilities, often due to suboptimal data mixtures and limited model parameter sizes. However, during our evaluation process with CS-Eval, we discovered that when continuing to train large language models, if the quality of the data is not high enough, it can even lead to a decrease in the model's overall domain capabilities. For example, SecGPT-13B is further trained from Baichuan-13B-Base, but we found that Baichuan-13B-Chat outperforms SecGPT-13B in terms of comprehensive effectiveness. Since the dataset used for further training of SecGPT-13B is open-source, we analyzed the reasons for the drop in comprehensive effectiveness.
\begin{itemize}
\item Pre-training Data Quality: In the open-source dataset for SecGPT-13B, we found that some data had not undergone rigorous quality filtering, which could have led to a decrease in model performance.
\item Instruction Fine-tuning Stage: We found that the model's ability to follow instructions weakened in the answers produced by SecGPT-13B. Therefore, during the instruction fine-tuning stage, the diversity of tasks, data quality, and the adequacy of training are crucial.
\end{itemize}
Therefore, in the training process of large language models, data quality and diversity are particularly critical. Multiple studies have indicated that data quality is sometimes more important than the sheer volume of data. For example, Qwen-Math, which underwent high-quality fine-tuning based on Qwen, achieved excellent performance in its domain.

\subsection{Impact of Model Parameter Size}
\begin{figure}[h]
\centering
\includegraphics[width=0.9\linewidth]{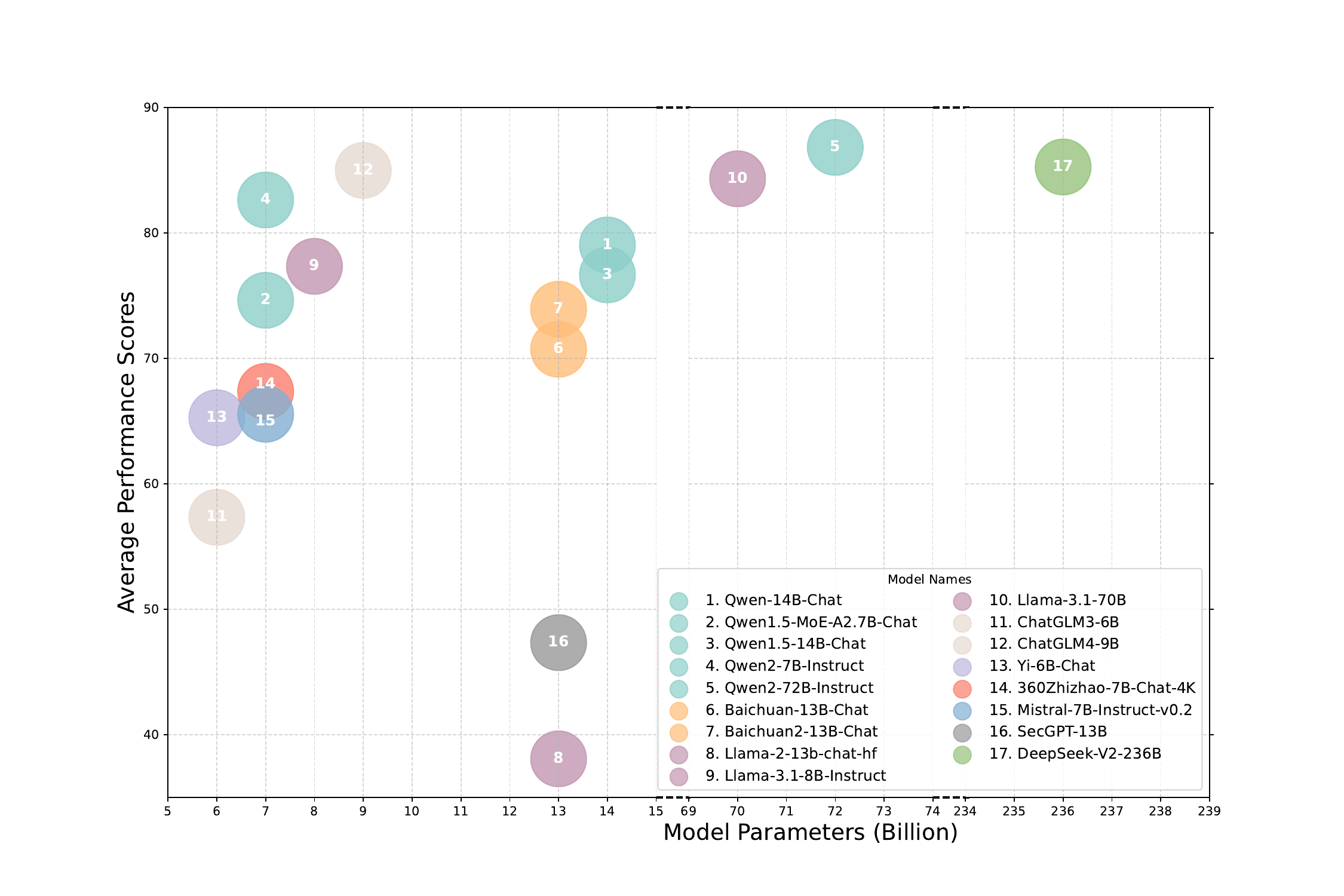}
\caption{The Average Scores of Models with Different Parameter Sizes.}
\label{fig:Parameter}
\end{figure}

Through analysis, we found that models with larger parameter sizes generally exhibit superior fine-grained knowledge differentiation, leading to better performance. For instance, as illustrated in Figure \ref{fig:Parameter}, both Llama-3.1-70B-Instruct compared to Llama-3.1-8B-Instruct, and Qwen2-72B-Instruct compared to Qwen2-7B-Instruct, clearly demonstrate this trend.

We found that in the MoE (Mixture of Experts) model during the inference phase, using approximately 20\% of the total parameters can achieve results close to using all parameters. Although there was a decline in performance in some domains, the overall score remained close. For example, Qwen1.5-MoE-A2.7B-Chat has a total parameter count of 14.3 billion, but only activates 2.7 billion parameters during the inference phase, yet the overall score is close to that of Qwen1.5-14B-Chat.

\subsection{Evolution of LLM Security Capabilities Over Time}
\begin{figure}[h]
\centering
\includegraphics[width=\linewidth]{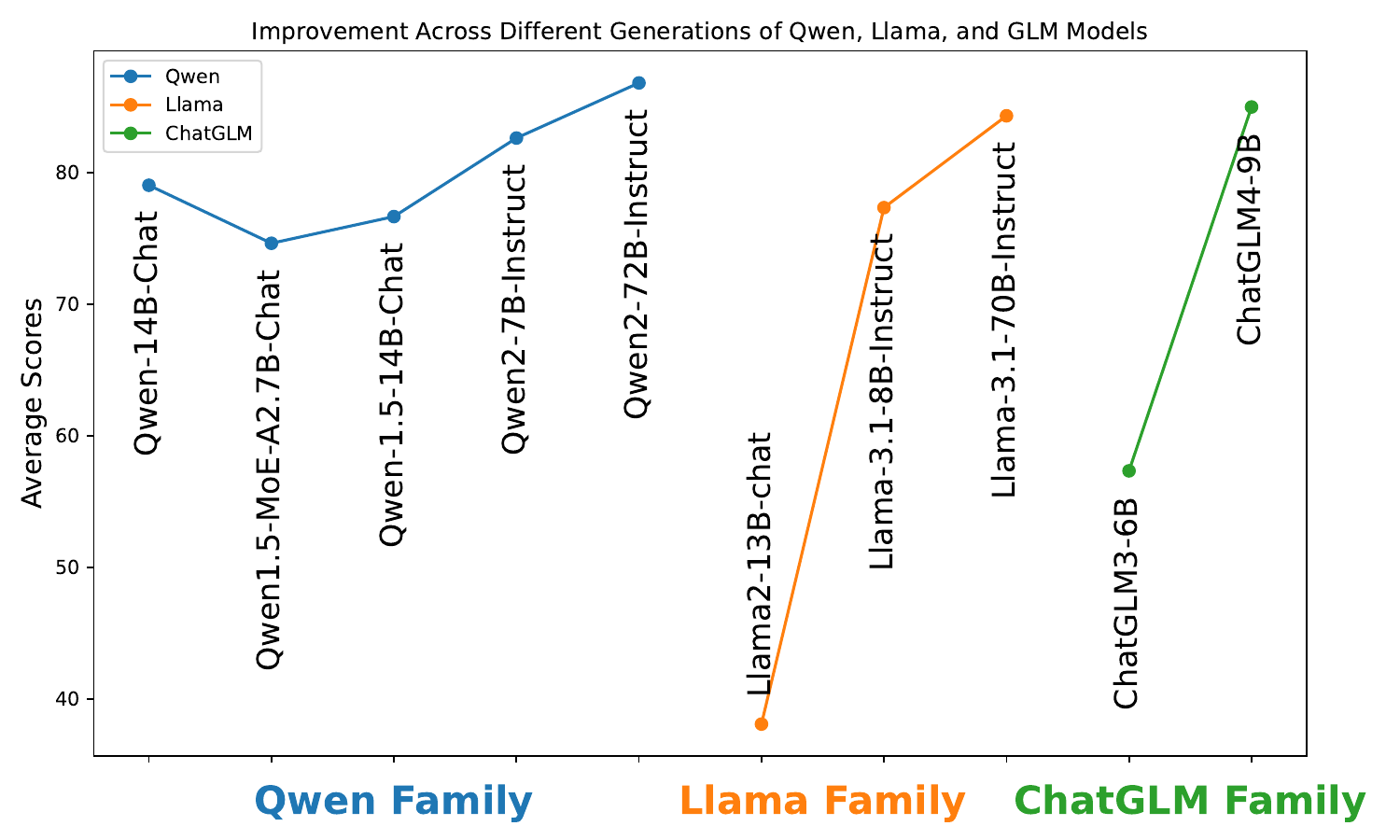}
\caption{Performance Scores Across Different Generations of Models.}
\label{fig:evolution}
\end{figure}

We conducted CS-Eval assessments in both May and August. Encouragingly, these LLMs exhibited significant improvements over this period, with some models showing notable score increases, as illustrated in Figure \ref{fig:evolution}. Based on our evaluation results at different time periods and the technical reports of large language models, we can conclude that with improvements in training data quality and strategies, large language models can perform better even with smaller model parameters. For instance, Qwen2-7B-Instruct outperforms Qwen-14B-Chat on overall score of CS-Eval. With the continuous iteration of foundational models, we can discern from their technical reports the importance of improving data quality. For example, Llama-3.1 and Qwen2 enhance data quality through methods such as source data augmentation, AUTOIF \cite{dong2024self}, and data cleaning. As a result, we found in our tests that models with the same parameter count performed better after being upgraded. For instance, Baichuan2-13B-Chat outperforms Baichuan-13B-Chat.

Interestingly, we found that newer, smaller LLMs can surpass older models with larger parameter counts, challenging the assumption that size always correlates with performance. We speculate that this is largely due to the amount and quality of training data. For example, Llama-3.1-8B-Instruct significantly outperforms Llama-2-13b-chat-hf.
Excitingly, as model capabilities improve, using synthetic data generated by higher-quality models for training also tends to yield performance gains \cite{yang2024qwen2,dubey2024llama}. For example, in our evaluation of Llama-3.1-8B-Instruct, a significant amount of synthetic data was introduced for training purposes.

\section{Discussion}

\subsection{Implication}
Through the analysis of our experiments, we found that the scaling law for parameter size is crucial, with larger models generally performing better. However, constructing high-quality, diverse datasets is equally important, especially when models are not fully trained, as data quality can sometimes outweigh the importance of parameter size. Additionally, open-source models are steadily approaching the performance levels of proprietary models, highlighting the strong potential of community-driven development.

\subsection{Limitations}
CS-Eval also has some limitations that need to be addressed for further enhancement.
Firstly, the reliance on manual data collection highlights the need to develop automated methods to reduce manual labor in the future.
Secondly, while the evaluation framework provides valuable insights, it could be refined to create a more integrated feedback loop that offers targeted guidance for model training and optimization.
Additionally, CS-Eval, though comprehensive, could expand into specialized areas like kernel exploitation, benefiting advanced cybersecurity research. We also plan to incorporate detailed code security assessments in the future, such as data flow and control flow analysis.
Lastly, we discovered that many cybersecurity tasks require specific execution environments, such as virtual setups for reproducing vulnerabilities. In the future, we plan to explore using agents to conduct these evaluations.

\section{Conclusion}
In this paper, we introduce CS-Eval, a comprehensive benchmark designed to assess the cybersecurity capabilities of large language models (LLMs). CS-Eval spans 11 categories and 42 subcategories, with a focus on critical tasks relevant to both industry and academia. And it evaluates models across three levels: knowledge, ability, and application.
The benchmark incorporates high-quality, manually constructed questions and supports dynamic benchmark data generation by LLMs. Through extensive experiments, we identified models that are particularly effective for specific cybersecurity tasks, alongside several noteworthy findings. We hope that CS-Eval can serve as an effective and continuously evolving benchmark, supporting the future development of LLMs' cybersecurity capabilities.

\bibliography{References}

\newpage
\appendix
\section{A Self-Instruct-Prompts}
\begin{figure}[h!]
\centering
\includegraphics[width=\linewidth]{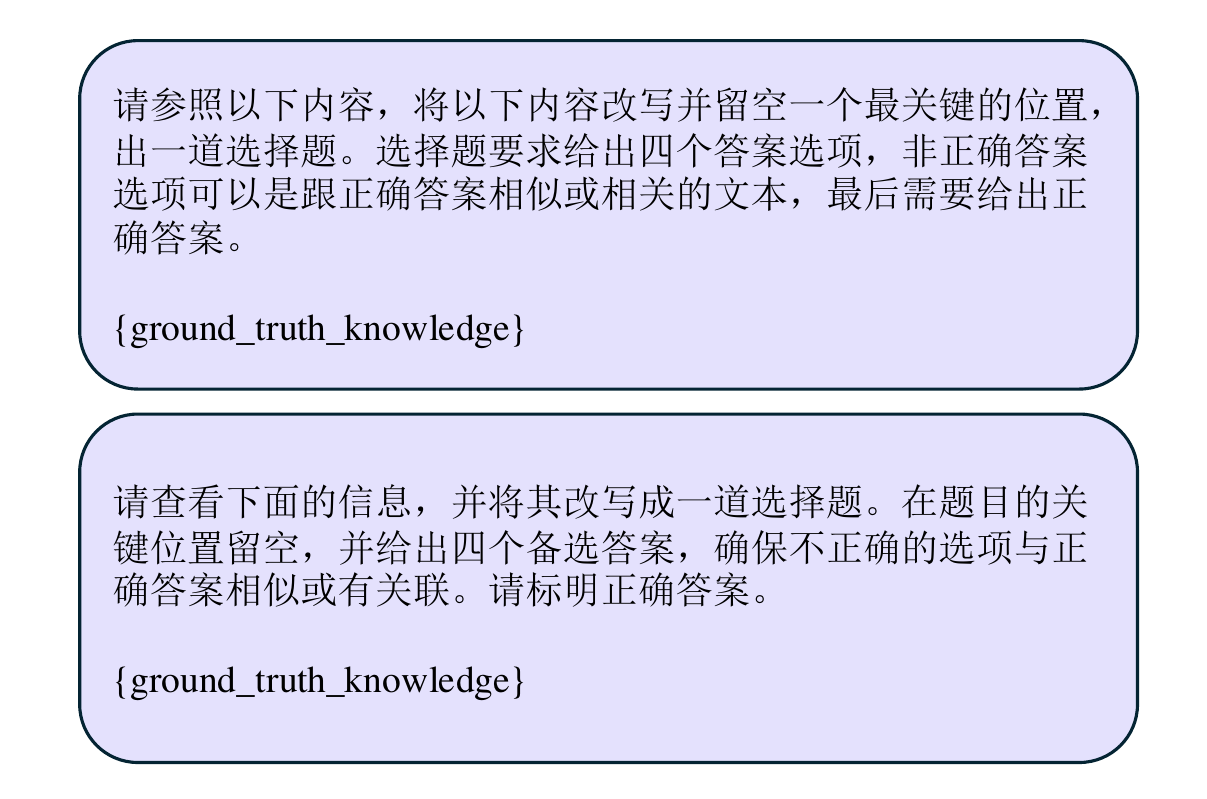}
\caption{Self-instruct prompt used for generating Chinese questions.}
\label{fig:prompt}
\end{figure}

The self-instruct prompts we utilized include the BingLingual version, as shown in Table \ref{table:instruct} and Figure \ref{fig:prompt}.
\begin{table}[h]
\centering
\begin{tabular}{|p{0.9\linewidth}|}
\hline
\textbf{Self-Instruct-Prompts} \\
\hline
1. Please refer to the following content and rewrite it to create a multiple-choice question, leaving one key position blank. The multiple-choice question should provide four answer options, with non-correct options being similar or related to the correct answer. The correct answer should be provided.\\
\{ground\_truth\_knowledge\}\\
\\
2. Please review the content below and reformulate it into a multiple-choice question, ensuring that one crucial part is left blank. Provide four possible answers, including the correct one, and make sure the incorrect choices are similar or related to the right answer.\\
\{ground\_truth\_knowledge\}\\
\\
3. Examine the following content and convert it into a multiple-choice question by leaving one key component blank. Offer four potential answers, making sure that the incorrect options are similar to the correct one. Indicate the correct answer.\\
\{ground\_truth\_knowledge\}\\
\\
4. Examine the following content and convert it into a true/false question. Offer yes or no answers.\\
\{ground\_truth\_knowledge\}\\
\hline
\end{tabular}
\caption{Instructional Prompts for Question Reformulation}
\label{table:instruct}
\end{table}

\section{B Cases of CS-Eval}
CS-Eval includes large, high-quality datasets. Below are some examples.
\begin{table}[h]
\centering
\begin{tabular}{|p{1\linewidth}|}
\hline
\textbf{Multiple-Choice Question:} \\
Suppose we have a user named Alice, and the following is her behavior timeline and activities: \\
\\
\textbf{Timeline:} \\
- 9:00 AM: Logged into the company network using a device she had never used before \\
- 9:30 AM: A large number of folders, including some sensitive files, were downloaded to her personal folder \\
- 10:00 AM: Executed some unauthorized system commands \\
- 11:00 AM: Multiple failed login attempts, and her account was locked \\
- 11:30 AM: After unlocking, she immediately performed a large number of file transfer operations \\
\\
What was the first sign of abnormal behavior for Alice's account? \\
A. Executed some unauthorized system commands \\
B. Multiple failed login attempts, and her account was locked \\
C. Logged into the company network using a device she had never used before \\
D. After unlocking, immediately performed a large number of file transfer operations \\
\\
Please select the single correct option. \\
\hline
\end{tabular}
\caption{Example Case 1: Alice's Behavior Timeline and Abnormal Activity}
\label{table:case1}
\end{table}

\begin{table}[h!]
\centering
\begin{tabular}{|lp{0.5\linewidth}c|}
\hline
\textbf{User/ IP Address} & \textbf{Behavior} & \textbf{Type} \\
\hline
Alice & File access and editing & xx \\
192.168.1.10 & Large file downloads & xx \\
Bob & Accessing internal company resources & xx \\
192.168.1.20 & Logged in during non-working hours and transferred a large amount of data & xx \\
Charlie & Website browsing activity & xx \\
David & Multiple attempts to log in with incorrect credentials & xx \\
192.168.1.30 & 5 data transfers within 1 minute & xx \\
\\
\multicolumn{3}{|p{0.9\linewidth}|}{\textbf{Question:} Based on the network behavior analysis, which user or IP address exhibited abnormal behavior?} \\
\multicolumn{3}{|p{0.9\linewidth}|}{A) Alice \quad B) 192.168.1.20 \quad C) Charlie \quad D) 192.168.1.30} \\
\hline
\end{tabular}
\caption{Example Case 2: Network Activity Analysis and Question}
\label{table:case2}
\end{table}

\begin{table}[h!]
\centering
\begin{tabular}{|p{\linewidth}|}
\hline
\textbf{Log:} \\
\texttt{2023-09-15 12:01:00 INFO [User: Alice] Login successful} \\
\texttt{2023-09-15 12:05:00 WARNING [User: Alice] Failed to access sensitive data} \\
\texttt{2023-09-15 12:10:00 INFO [User: Bob] Login successful} \\
\texttt{2023-09-15 12:15:00 INFO [User: Bob] Accessed salary.docx} \\
\texttt{2023-09-15 12:20:00 INFO [User: Charlie] Login successful} \\
\texttt{2023-09-15 12:25:00 WARNING [User: Charlie] Multiple failed login attempts} \\
\texttt{2023-09-15 12:30:00 INFO [User: Alice] Accessed confidential.docx} \\
\texttt{2023-09-15 12:35:00 INFO [User: Bob] Login successful} \\
\texttt{2023-09-15 12:40:00 INFO [User: Bob] Accessed project\_plan.docx} \\
\texttt{2023-09-15 12:45:00 WARNING [User: Alice] Unauthorized attempt to access sensitive data} \\
\texttt{2023-09-15 12:45:05 INFO [User: Alice] Accessed home page} \\
\texttt{2023-09-15 12:45:08 INFO [User: Alice] Accessed "Friends" page} \\
\texttt{2023-09-15 12:45:08 INFO [User: Alice] called sendMessage API} \\
\texttt{2023-09-15 12:46:00 INFO [User: Bob] called sendMessage API} \\
\texttt{2023-09-15 12:46:00 INFO [User: Bob] called sendMessage API} \\
\texttt{2023-09-15 12:46:00 INFO [User: Bob] called sendMessage API} \\
\texttt{2023-09-15 12:46:00 INFO [User: Bob] called sendMessage API} \\
\texttt{2023-09-15 12:46:00 INFO [User: Bob] called sendMessage API} \\
\texttt{2023-09-15 12:46:00 INFO [User: Bob] called sendMessage API} \\
\texttt{2023-09-15 12:46:00 INFO [User: Bob] called sendMessage API} \\
\texttt{2023-09-15 12:46:01 INFO [User: Bob] called sendMessage API} \\
\texttt{2023-09-15 12:46:01 INFO [User: Bob] called sendMessage API} \\
\texttt{2023-09-15 12:46:01 INFO [User: Bob] called sendMessage API} \\
\texttt{2023-09-15 12:46:02 INFO [User: Bob] called sendMessage API} \\
\\
\textbf{Question:} \\
Based on the log records, which user's behavior most likely requires further investigation and review, and what is the corresponding reason? \\
A) Alice, attempting to access sensitive data without authorization \\
B) Bob, sending messages directly through the API without following the normal process on the friends page, indicating potential data misuse or malicious behavior, requiring further investigation and review \\
C) Charlie, multiple failed login attempts, which could indicate an automated testing tool being obstructed at the login page, potentially under the control of a malicious group \\
D) None of the above \\
\hline
\end{tabular}
\caption{Example Case 3: Log Analysis Question}
\label{table:case3}
\end{table}

\clearpage
\section{C Prompt of Rewrite the Questions of CS-Eval to Generating Dynamic Data}
\label{chap:appendix_c}
This section introduces the prompts used for dynamic data generation. We employ two strategies: the first is detailed in Table \ref{table:dynamic1}, and the second is outlined in Table \ref{tab:dynamic2}.
\begin{table}[h!]
\centering
\begin{tabular}{|p{0.9\linewidth}|}
\hline
\textbf{Prompts for Dynamic Question Generation, Part 1}\\
\hline
1. Rewrite the following question by changing the subject from 'administrator' to 'user'. Ensure the question still tests the same core concept. \\
\{Original Question\}\\
\\
2. Rewrite the following multiple-choice question by modifying the options to include alternative technologies or methods that still test the same concept. \\
\{Original Question\}\\
\\
3. Rewrite the following question by changing the scenario to a different environment, such as an e-commerce website. Ensure the question remains relevant to the original cybersecurity concept. \\
\{Original Question\}\\
\\
4. Reframe the following question by adding specific background information related to a banking application. Ensure the question tests the same concept. \\
\{Original Question\}\\
\\
5. Modify the following question by adding misleading or extraneous information to make it more challenging, but ensure the correct answer is still clear. \\
\{Original Question\}\\
\\
6. Reverse Questioning: By posing the question in a reversed or opposite manner, you assess whether the examinee's understanding of the concept is comprehensive and adaptable. For example, if the original question asks about the result of a specific action, the reversed question could ask about the conditions required for that result to occur. \\
\{Original Question\}\\
\\
7. Expand the following question into a multi-step problem that requires the respondent to perform several related tasks or analyses. \\
\{Original Question\}\\
\\
8. Restructure the following question to change the logical flow, possibly by rearranging the order of the components or by introducing sub-questions. \\
\{Original Question\}\\
\\
9. Create a scenario-based question from the following by adding a realistic context or story, requiring the respondent to analyze the situation and make decisions. \\
\{Original Question\}\\
\hline
\end{tabular}
\caption{Prompts for Question Reformulation}
\label{table:dynamic1}
\end{table}

\begin{table}[h!]
\centering
\begin{tabular}{|p{0.9\linewidth}|}
\hline
\textbf{Prompts for Dynamic Question Generation, Part 2} \\
\hline
\textbf{1. Summarize Knowledge Point} \\
Based on the given questions, please summarize the knowledge points being tested behind these topics. \\
The questions are: \{questions\} \\
Reply to me in the following format: \\
\texttt{```json} \\
\texttt{['knowledge point 1', 'knowledge point 2', ......]} \\
\texttt{```} \\
\\
\textbf{2. Generate Questions} \\
Generate 1 challenging question about \{points\}. \\
Then rewrite it as a multiple-choice question by leaving one key position blank. The multiple-choice question should offer four answer options, with the incorrect options being similar or related to the correct one. Provide the correct answer as well. Reply in the following format: \\
\texttt{```json} \\
\texttt{['question 1']} \\
\texttt{```} \\
\hline
\end{tabular}
\caption{Prompts for Dynamic Question Generation}
\label{tab:dynamic2}
\end{table}

\clearpage
\section{D Complete Experimental Results}
The complete set of experiments evaluating all large language models and covering all designed fields of cybersecurity is presented in Table \ref{table:all_combined}.
\begin{table}[h!]
\centering
\resizebox{\textwidth}{!}{
\begin{tabular}{cccccccc}
\hline
\multicolumn{1}{c}{Model} & \multicolumn{1}{c}{Business Continuity} & \multicolumn{1}{c}{Security Architecture} & \multicolumn{1}{c}{Supply Chain} & \multicolumn{1}{c}{Data Security and} & \multicolumn{1}{c}{Threat Detection} & \multicolumn{1}{c}{Vulnerability Management} & \multicolumn{1}{c}{AI and Cybersecurity} \\
\multicolumn{1}{c}{} & \multicolumn{1}{c}{and Emergency Response} & \multicolumn{1}{c}{Design} & \multicolumn{1}{c}{Security} & \multicolumn{1}{c}{Privacy Protection} & \multicolumn{1}{c}{and Prevention} & \multicolumn{1}{c}{and Penetration Testing} & \multicolumn{1}{c}{} \\
\hline
GPT4-8K                  & 84.28 & 83.90  & 89.30 & 86.90     & 85.21   & 89.63     & 91.58   \\
Qwen2-72B-Instruct       & 77.33 & 85.37  & 86.38 & 86.31     & 88.56   & 88.19     & 86.63   \\
GPT4o                    & 77.00 & 83.41  & 89.70 & 84.92     & 87.82   & 85.44     & 89.60   \\
DeepSeek-V2-0628         & 78.33 & 83.90  & 89.37 & 83.33     & 84.87   & 85.60     & 89.60   \\
ChatGLM4                 & 79.67 & 82.44  & 90.70 & 83.73     & 84.13   & 85.44     & 88.61   \\
Llama-3.1-70B-Instruct   & 74.33 & 79.02  & 86.71 & 86.11     & 85.61   & 84.95     & 86.14   \\
Qwen2-7B-Instruct        & 73.00 & 80.00  & 87.38 & 79.56     & 84.32   & 82.52     & 81.19   \\
GPT3.5-Turbo-16K         & 81.27 & 76.59  & 88.96 & 82.14     & 79.52   & 80.71     & 80.69   \\
Qwen-14B-Chat            & 78.60 & 74.15  & 87.63 & 76.68     & 79.67   & 77.80     & 87.13   \\
Llama-3.1-8B-Instruct    & 71.67 & 76.59  & 85.05 & 75.99     & 76.57   & 77.99     & 79.70   \\
Qwen1.5-14B-Chat         & 70.23 & 70.73  & 81.27 & 77.58     & 77.53   & 75.77     & 78.71   \\
Qwen1.5-MoE-A2.7B-Chat   & 72.24 & 68.78  & 81.94 & 70.24     & 76.61   & 74.80     & 74.75   \\
Baichuan2-13B-Chat       & 73.91 & 71.71  & 80.27 & 75.69     & 76.94   & 70.55     & 76.24   \\
Baichuan-13B-Chat        & 69.00 & 70.73  & 75.42 & 71.43     & 69.37   & 71.04     & 76.73   \\
360Zhinao-7B-Chat-4K     & 66.33 & 68.78  & 70.00 & 65.02     & 68.63   & 64.78     & 71.29   \\
Mistral-7B-Instruct-v0.2 & 63.67 & 62.44  & 72.76 & 63.44     & 64.40   & 63.71     & 69.31   \\
Yi-6B-Chat               & 59.67 & 60.00  & 72.76 & 62.85     & 63.85   & 64.68     & 65.84   \\
ChatGLM3-6B              & 56.67 & 61.46  & 68.44 & 57.71     & 61.47   & 50.81     & 65.35   \\
SecGPT-13B               & 45.33 & 45.85  & 59.14 & 43.08     & 47.34   & 46.77     & 40.59   \\
Llama-2-13b-chat-hf      & 39.13 & 37.07  & 30.43 & 35.52     & 39.00   & 38.57     & 38.12   \\
\hline
\end{tabular}}

\vspace{0.3cm}

\resizebox{\textwidth}{!}{
\begin{tabular}{cccccccc}
\hline
\multicolumn{1}{c}{Model} & \multicolumn{1}{c}{Infrastructure} & \multicolumn{1}{c}{Encryption Technology} & \multicolumn{1}{c}{Access Control and} & \multicolumn{1}{c}{System Security and} & \multicolumn{1}{c}{Chinese} & \multicolumn{1}{c}{English} & \multicolumn{1}{c}{Average} \\
\multicolumn{1}{c}{} & \multicolumn{1}{c}{Security} & \multicolumn{1}{c}{and Key Management} & \multicolumn{1}{c}{Identity Management} & \multicolumn{1}{c}{Software Security Fundamentals} & \multicolumn{1}{c}{} & \multicolumn{1}{c}{} & \multicolumn{1}{c}{Scores} \\
\hline
GPT4-8K                  & 88.83  & 86.51     & 86.56     & 90.00       & 87.96   & 82.19   & 87.57 \\
Qwen2-72B-Instruct       & 87.69  & 88.32     & 88.70     & 86.33       & 87.05   & 83.56   & 86.82 \\
GPT4o                    & 87.35  & 88.32     & 84.29     & 90.00       & 85.95   & 88.36   & 86.11 \\
DeepSeek-V2-0628         & 88.19  & 83.94     & 83.33     & 88.00       & 85.23   & 85.62   & 85.26 \\
ChatGLM4-9B              & 87.19  & 83.21     & 84.67     & 84.33       & 85.26   & 81.16   & 84.99 \\
Llama-3.1-70B-Instruct   & 86.02  & 85.04     & 82.76     & 86.33       & 84.08   & 87.67   & 84.32 \\
Qwen2-7B-Instruct        & 82.86  & 87.59     & 85.06     & 83.33       & 82.82   & 79.79   & 82.63 \\
GPT3.5-Turbo-16K         & 83.17  & 69.59     & 78.31     & 80.00       & 80.62   & 80.14   & 80.59 \\
Qwen-14B-Chat            & 81.33  & 68.49     & 78.89     & 77.00       & 79.99   & 65.41   & 79.04 \\
Llama-3.1-8B-Instruct    & 80.37  & 69.71     & 76.82     & 78.33       & 77.14   & 80.14   & 77.34 \\
Qwen1.5-14B-Chat         & 78.00  & 76.13     & 77.59     & 75.33       & 76.68   & 75.68   & 76.66 \\
Qwen1.5-MoE-A2.7B-Chat   & 71.88  & 73.50     & 79.50     & 74.33       & 75.99   & 55.14   & 74.63 \\
Baichuan2-13B-Chat       & 76.50  & 60.09     & 73.90     & 70.67       & 73.79   & 75.34   & 73.92 \\
Baichuan-13B-Chat        & 74.71  & 52.92     & 73.18     & 68.33       & 70.47   & 74.32   & 70.73 \\
360Zhinao-7B-Chat-4K     & 66.78  & 51.04     & 68.14     & 67.67       & 66.68   & 61.99   & 66.37 \\
Mistral-7B-Instruct-v0.2 & 70.43  & 57.78     & 69.54     & 63.67       & 66.01   & 63.36   & 65.93 \\
Yi-6B-Chat               & 64.84  & 68.80     & 69.98     & 63.00       & 65.58   & 59.93   & 65.27 \\
ChatGLM3-6B              & 59.87  & 47.78     & 55.26     & 50.33       & 57.14   & 59.25   & 57.33 \\
SecGPT-13B               & 47.60  & 41.54     & 53.15     & 46.00       & 48.45   & 31.85   & 47.34 \\
Llama-2-13b-chat-hf      & 37.67  & 34.11     & 47.60     & 33.33       & 38.40   & 32.88   & 38.08 \\
\hline
\end{tabular}}
\caption{Experimental Results}
\label{table:all_combined}
\end{table}

\end{document}